\documentstyle[11pt,aaspp4]{article}

\font\caps=cmcsc10 scaled 1200
\def\etal {et~al.}
\def\radm2{radians m$^{-2}$}

\def\HI {H\kern0.1em{\sc i}}
\def\pks {PKS\,2322$-$123}

\slugcomment{submitted to the Astrophysical Journal Letters}

\lefthead{Taylor et al.}
\righthead{\HI\ in the Nucleus of PKS\,2322$-$123}

\begin{document}

\title{~~\\ ~~\\ \HI\ Absorption Toward the Nucleus of the Radio
  Galaxy \\ \pks\ in A\,2597}

\author{G.~B.~Taylor\altaffilmark{1}, C.~P.~O'Dea\altaffilmark{2}, A.~B.~Peck\altaffilmark{1,3}, \& A.~M.~Koekemoer\altaffilmark{2}}

\altaffiltext{1}{National Radio Astronomy Observatory, Socorro, NM
  87801, USA; gtaylor@nrao.edu, apeck@nrao.edu}
\altaffiltext{2}{Space Telescope Science Institute, Baltimore, MD
  21218, USA; odea@stsci.edu, koekemoe@stsci.edu}
\altaffiltext{3}{New Mexico Institute of Mining and Technology,
  Socorro, NM 87801, USA}

\begin{abstract}
  We present sensitive, high resolution Very Long Baseline Array
  (VLBA) observations of the central 0.3\arcsec\ of \pks\ at 1.3 and 5
  GHz.  These observations reveal straight and symmetric jets emerging
  from both sides of an inverted spectrum core. The 21 cm line of
  atomic hydrogen is detected in absorption against the core and
  eastern jet with substantial opacities, but is not seen towards the
  equally strong western jet.  Both a narrow (110 km s$^{-1}$ FWHM)
  and a very broad (735 km s$^{-1}$ FWHM) line are seen, although the
  very broad line is seen only against the core.  Both lines are
  redshifted ($\sim$220 $\pm$ 100 km s$^{-1}$) with respect to the
  systemic velocity. The most likely explanation for the observed \HI\ 
  kinematics are an atomic torus centered on the nucleus with
  considerable turbulence and inward streaming motions. The scale
  height of this torus is less than 20 pc.  Though rare in flux
  limited samples of compact radio sources, symmetric parsec-scale
  structure appears nearly ubiquitous among radio galaxies with \HI\ 
  absorption, probably because they are viewed more nearly edge-on
  through the torus.
\end{abstract}


\vspace{2cm}

\section{Introduction}

\HI\ gas has now been detected towards several radio galaxies embedded
in X-ray cooling flow clusters: 3C\,84 in A\,426 (Crane \etal\ 1982;
Jaffe 1990); Hydra~A in A\,780 (Taylor 1996; Dwarakanath, Owen, \& van
Gorkom 1995), and recently \pks\ in A\,2597 (O'Dea, Baum \& Gallimore
1994).  In an even larger number of such radio galaxies \HI\ has not
been detected in spite of fairly intensive searches (e.g. Virgo~A, and
3C~338 in A\,2199 -- Dwarkanath, van Gorkom, \& Owen 1994; Jaffe 1991;
O'Dea, Gallimore, \& Baum 1995).  Besides a natural curiosity about
these unusual systems, additional motivation to search for \HI\ has
come from the discovery of excess low-energy X-ray absorption (White
\etal\ 1991) which has been suggested to be due to a population of
cold \HI\ clouds in the cluster with a column density of N$_{\rm H}$
$\sim 10^{21}$ cm$^{-2}$ and covering factor near unity.  To date this
population has not been seen and the \HI\ gas that is detected seems
to be closely associated with the active nucleus.  In Hydra~A the \HI\ 
gas appears to be distributed in a disk having a rotation axis closely
aligned with the jet axis and with a height of just 30 pc.

O'Dea, Baum \& Gallimore (1994) detected two \HI\ components in PKS
2322$-$123 using the VLA in its A configuration.  One component is a
narrower (220 km s$^{-1}$ wide) feature seen only towards the center of the
source (the location of the nucleus).  The other component is broader
(410 km s$^{-1}$ wide) and spatially extended against the entire radio
source.  O'Dea \etal\ suggest that the ``broad'' component is
associated with a bright H$\alpha$ nebula.

A detailed study of the optical spectrum from the central
$\sim$kiloparsec of A\,2597 by Voit \& Donahue (1997) allowed them to
measure the density, temperature and metal abundances of the
line-emitting gas and conclude that the most likely explanation for
the brilliant ionizing nebula in A\,2597 is a population of hot stars.
As a byproduct of their study they determine the redshift of \pks\ to
be 0.0821 $\pm$ 0.0001.  HST observations by Koekemoer \etal\ (1998)
reveal a series of blue knots and clumps that are most likely regions
of recent star formation.  They also find emission line filaments at
the edges of the radio lobes, which argues for a direct interaction
between the radio source and its environment.  Falcke \etal\ (1998)
recently detected molecular hydrogen in the central kiloparsecs of
\pks.  They suggest that the luminous H$_2$ emission correlates with
the presence of a strong radio source, luminous H$\alpha$ emission and
a cooling flow of $>$100 M$_\odot$ y$^{-1}$.

We assume H$_0 = 75$ km s$^{-1}$ Mpc$^{-1}$ and q$_0$ = 0 throughout.

\section{VLBA Observations}

The VLBA observations were carried out at 1.312 and 4.991 GHz with
the 10-element VLBA and 27-element VLA of the NRAO\footnote{The
National Radio Astronomy Observatory is operated by Associated
Universities, Inc., under cooperative agreement with the National
Science Foundation} in a single 10.5-hour observing session on 1996
December 7. The VLA was operated in ``phased-array'' mode to
synthesize a single aperture with a large collecting area.  The
calibrator J2246-1206 was used to phase up the VLA every 25 minutes.
The net integration time on \pks\ was 323 minutes at 1.3 GHz and 95
minutes at 5 GHz spread over multiple 19 minute snapshots to improve
u,v coverage.  Both right- and left-circular polarizations were
observed with 2-bit sampling.  The 1.3 and 5 GHz data were correlated
in separate passes so as to provide high spectral resolution (1024
channels per 8 MHz wide IF band) without the RL and LR
cross-correlation products at 1.3 GHz, and low spectral resolution (16
channels per IF band) with full cross-correlation products at 5 GHz.

Standard a priori flagging, amplitude calibration, fringe fitting,
bandpass calibration (using 3C\,84), and frequency averaging
procedures were followed in AIPS.  Global fringe fitting was performed
on J2246-1206 and the resulting delays and rates were transfered to
\pks.  Furthermore, after imaging, J2246-1206 was used in
conjunction with CALIB to obtain time-dependent amplitude corrections
which were also applied to A\,2597.  Subsequently, A\,2597 was phase-only
self-calibrated with a 1 minute solution interval. No amplitude
self-calibration was performed on A\,2597 itself.  All manual editing,
imaging, deconvolution, and self-calibration were performed using
{\caps Difmap} (Shepherd, Pearson, \& Taylor 1994, 1995).  Polarization
calibration was performed on the 5 GHz data using J2246-1206 to
determine the instrumental leakage terms, and the absolute polarization
angle calibration was determined from a single 12 minute observation
of 3C\,286 which has known polarized structure (Cotton \etal\ 1997).


\section{Continuum Results}

In Fig.~1 we show the naturally weighted 1.3 and 5 GHz VLBA images.
Both images display a clearly symmetric structure along a position
angle of $\sim$70$^\circ$.  On kiloparsec scales a bright hot spot
0.5\arcsec\ south-west of the nucleus (Sarazin \etal\ 1995) suggests
some bending of the jet to a position angle of $\sim$50$^\circ$.  At
the hot spot the south-west radio jet appears to be sharply
deflected to the south, giving the large scale structure an overall
``C'' shape.  

In the parsec scale jet at 1.3 GHz there are 5 components extended
nearly in a line over 63 mas (89 pc) with the strongest component at
the center.  These components are labeled in Fig.~1, and parameterized
by Gaussian modelfitting in Table 1.  The central component is
considerably stronger at 5 GHz and has a spectral index, $\alpha$, of
0.6 $\pm$ 0.1 where $S_\nu \propto \nu^\alpha$.  We identify this
component as the core based on its inverted spectrum, compactness,
strength, and the symmetric morphology of the source.  No significant
polarized flux density was measured at 5 GHz above the noise floor of
40 $\mu$Jy/beam.  This indicates a fractional polarization of $<$0.1\%
for the nucleus, and $<$10\% for components E2 and W2.

\begin{figure}
\vspace{16cm}
\includegraphics{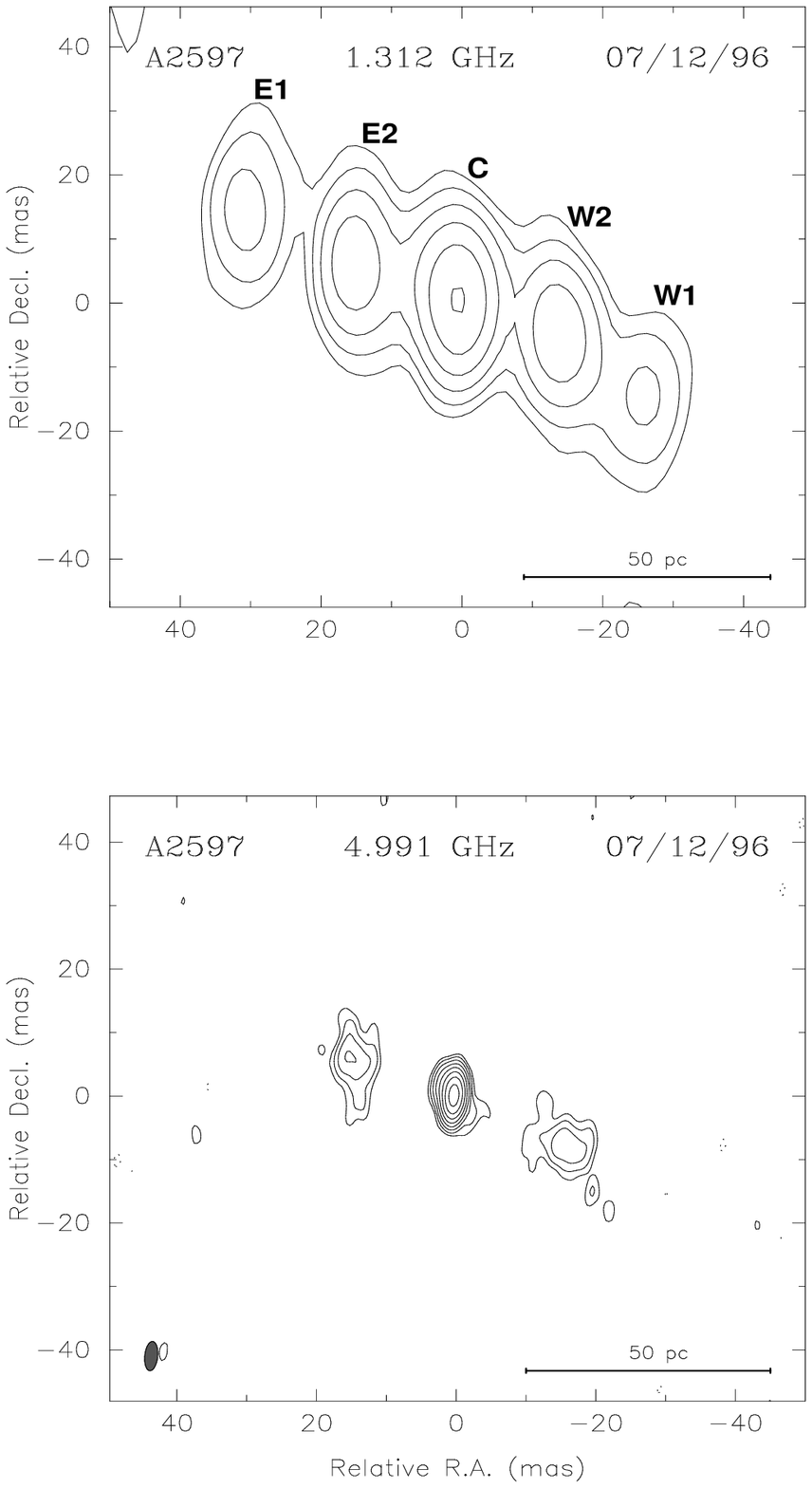}
\caption{The naturally weighted images of \pks\ at 1.3 and 5 GHz. The
restoring beam has dimensions 16.3 $\times$ 6.7 mas in position angle
$-$1.3$^\circ$ at 1.3 GHz and 4.63 $\times$ 1.86 mas in position angle
$-$5.1$^\circ$ at 5 GHz. Contours are at factor 2 intervals and start
at 0.5 and 0.2 mJy/beam for the 1.3 and 5 GHz images respectively.
\label{fig1}}
\end{figure}

\placefigure{fig1}

\begin{center}
TABLE 1 \\
\smallskip
C{\sc omponent} F{\sc lux} D{\sc ensities} {\sc and} S{\sc pectra}
\smallskip

\begin{tabular}{l r r r r r r r r r}
\hline
\hline
Component     & $r$ & $\theta$ &  $S_{\rm 1.3 GHz}$   &  $S_{\rm 5 GHz}$   & $\alpha_{1.3-5}$ \\
 & (mas) & ($^\circ$) & (mJy) & (mJy) &  \\
\hline
\noalign{\vskip2pt}
E1 & 33  &  64 & 2.8 $\pm$ 0.5 & $<0.2$ &  $<-1.9$ \\
E2 & 16  &  67 & 6.7 $\pm$ 0.6 &  4.8 $\pm$ 0.3 & $-$0.25 $\pm$ 0.1 \\
Core & 0 &  0 & 16.6 $\pm$ 0.9 & 36.9 $\pm$ 1.9 & $+$0.60 $\pm$ 0.1 \\
W2 & 16  &  $-$112 & 7.6 $\pm$ 0.6 &  4.1 $\pm$ 0.3 & $-$0.47 $\pm$ 0.1 \\
W1 & 30  &  $-$115 & 2.4 $\pm$ 0.5 & $<0.2$ & $<-1.9$  \\
\hline
\label{tab1}\end{tabular}
\end{center}

A spectral index map of the nucleus of \pks, made by tapering the 5
GHz image to the 1.3 GHz resolution, is shown in Fig.~2.  The two
images were aligned using the relatively steep spectrum component W2,
rather than the core, because if the core component is the location in
the jet where the synchrotron self absorption optical depth is unity,
then its position will vary with frequency (Blandford \& K\"onigl
1979), such that higher frequencies look ``deeper'' down the jet
towards the true center of activity.  Also, a gradient in the
free-free opacity towards the core may result in a shift in the
apparent core location as a function of frequency.  Given the low 5
GHz brightness temperatures for E2, C and W2 of 1.4 $\times 10^4$,
$\sim10^7$, and 1.2 $\times 10^4$ K respectively, synchrotron self
absorption is unlikely to be important.  Higher frequency VLBI observations
are needed to determine the intrinsic flux density and spectral index
before the free-free optical depth, $\tau_{ff}$, can be measured, but a
rough estimate on $\tau_{ff}$ can be obtained by extrapolating the
5 GHz flux density to 1.3 GHz assuming a spectral index of $-$1.  This
gives $\tau_{ff}$ of 1, 2.1 and 0.7 for E2, C and W2 respectively.  
These optical depths require emission measures of 7 $\times 10^7$,
14 $\times 10^7$, and 5 $\times 10^7$ cm$^{-3}$ pc for E2, C and W2
assuming a gas temperature of 10$^4$ K (see Levinson, Laor, \&
Vermeulen 1995).

\begin{figure}
\vspace{16cm}
\includegraphics{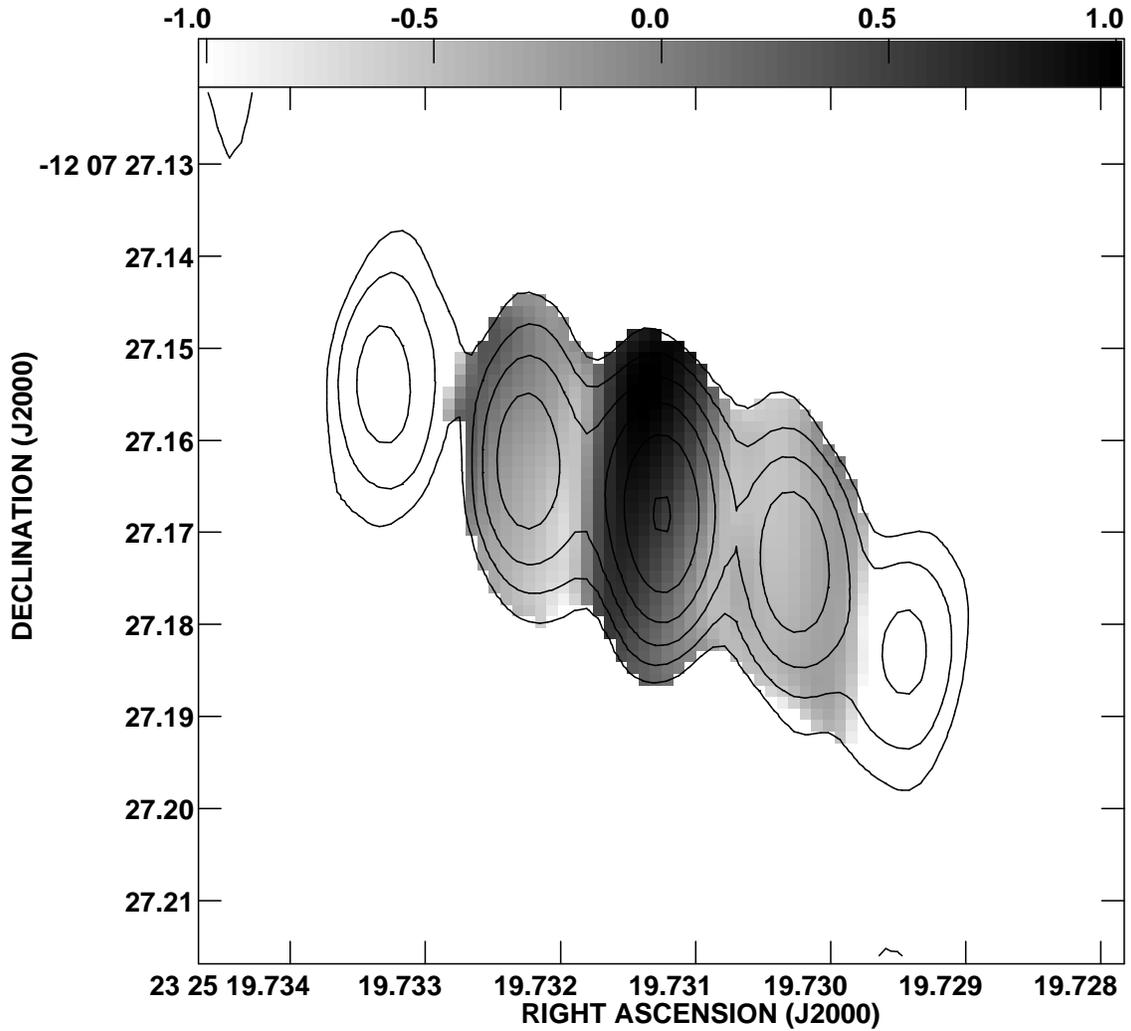}
\caption{A spectral index map of \pks\ between 1.3 and 5 GHz with 
1.3 GHz contours overlaid.  The 5 GHz image
has been heavily tapered and restored to the same resolution 
as the 1.3 GHz image.  The images have been registered using 
component W2.
\label{fig2}}
\end{figure}

\placefigure{fig2}

\section{\HI\ Absorption Results}

In Fig.~3 we show the 1.3 GHz continuum emission and the spectra of
each of the 5 components.  Both a very broad (735 km s$^{-1}$) and a
narrow (106 km s$^{-1}$) line are seen against the core component.
Only the narrow line is seen against component E2, and neither line is
detected towards W2. Measurements of the lines are given in Table 2
and shown graphically in Fig.~4.  Both of these lines are
significantly redshifted with respect to the velocity centroid of the
optical emission lines from the nucleus.  Voit \& Donahue (1997) have
obtained a high signal-to-noise 

\begin{center}
TABLE 2 \\
\smallskip
H{\sc I} C{\sc olumn} D{\sc ensities}
\smallskip

\begin{tabular}{l r r r r r r r r r}
\hline
\hline
Component &  $S_{\rm 1.3 GHz}$  & $S_{\rm line}$ & $V$ & $\Delta V$ & $\tau$ & $N_{\rm H}$ \\
(1) & (2) & (3) & (4) &  (5) & (6) & (7) \\
\hline
\noalign{\vskip2pt}
E1 -- narrow & 2.8 $\pm$ 0.5 & $<$ 1.5 mJy & -- & -- & $<$ 0.8 & -- \\
E2 -- narrow  & 6.7 $\pm$ 0.6 &  3.1 $\pm$ 0.3 & 24882 $\pm$ 6\phantom{0} & 133 $\pm$ 14 & 0.63 & 12.2 $\times$ 10$^{23}$ \\
Core -- narrow & 16.6 $\pm$ 0.9 & 2.3 $\pm$ 0.6 & 24880 $\pm$ 12 & 106 $\pm$ 33 & 0.15 & 2.3 $\times$ 10$^{23}$ \\
W2 -- narrow & 7.6 $\pm$ 0.6 & $<$ 2.0 mJy & -- & -- & $<$ 0.3 & -- \\
W1 -- narrow & 2.4 $\pm$ 0.5 & $<$ 1.5 mJy & -- & -- & $<$ 1.0 & -- \\
\\
E1 -- broad & 2.8 $\pm$ 0.5 & $<$ 1.0 mJy & -- & -- & $<$ 0.5 & -- \\
E2 -- broad  & 6.7 $\pm$ 0.6 & $<$ 1.0 mJy & -- & -- & $<$ 0.2 & -- \\
Core -- broad & 16.6 $\pm$ 0.9 & 3.8 $\pm$ 0.3 & 24905 $\pm$ 19 & 735 $\pm$ 53 & 0.26 & 27.8 $\times$ 10$^{23}$ \\
W2 -- broad & 7.6 $\pm$ 0.6 & $<$ 1.0 mJy & -- & -- & $<$ 0.2 & -- \\
W1 -- broad & 2.4 $\pm$ 0.5 & $<$ 1.0 mJy & -- & -- & $<$ 0.5 & -- \\
\hline
\label{tab2}\end{tabular}
\end{center}
\begin{center}
{\sc Notes to Table 2}
\end{center}
\noindent
Col.(1).---Component name and type of \HI\ line detected (or upper limit
given for).
Col.(2).---Continuum flux density at 1.3 GHz in mJy.
Col.(3).---Depth of the line in mJy or 3$\sigma$ upper limit.
Col.(4).---Central velocity in km s$^{-1}$.
Col.(5).---FWHM in velocity in km s$^{-1}$.
Col.(6).---Optical depth or 3$\sigma$ upper limit.
Col.(7).---The column density in units of cm$^{-2}$ calculated assuming
a covering factor of 1 and a spin temperature of 8000 K (Conway \&
Blanco 1995).
\bigskip

\noindent
spectrum from which they derive an
emission line redshift of 24613 $\pm$ 60 km s$^{-1}$. The presence of
large scale asymmetries in the emission line gas kinematics, however,
may indicate that the gas is substantially offset from the systemic
velocity.  Stellar absorption features may better trace a more relaxed
inner potential.  Taking advantage of prominent Ca~II H, K, and G band
absorption features we performed a cross correlation analysis of
the Voit \& Donahue spectrum with a range of galaxy template spectra.
Correlations with different spectra gave consistent results with a
mean systemic velocity of 24673 $\pm$ 20 km s$^{-1}$.  Given the
peculiar nature of \pks\ (elliptical with strong emission lines),
which is not well matched to any of the templates, and the mismatch in
spectral resolution, a more realistic error is $\pm$100 km s$^{-1}$.
The mean offset of the \HI\ absorption features from the systemic
velocity is thus 220 $\pm$ 100 km s$^{-1}$.

\begin{figure}
\vspace{16.5cm}
\includegraphics{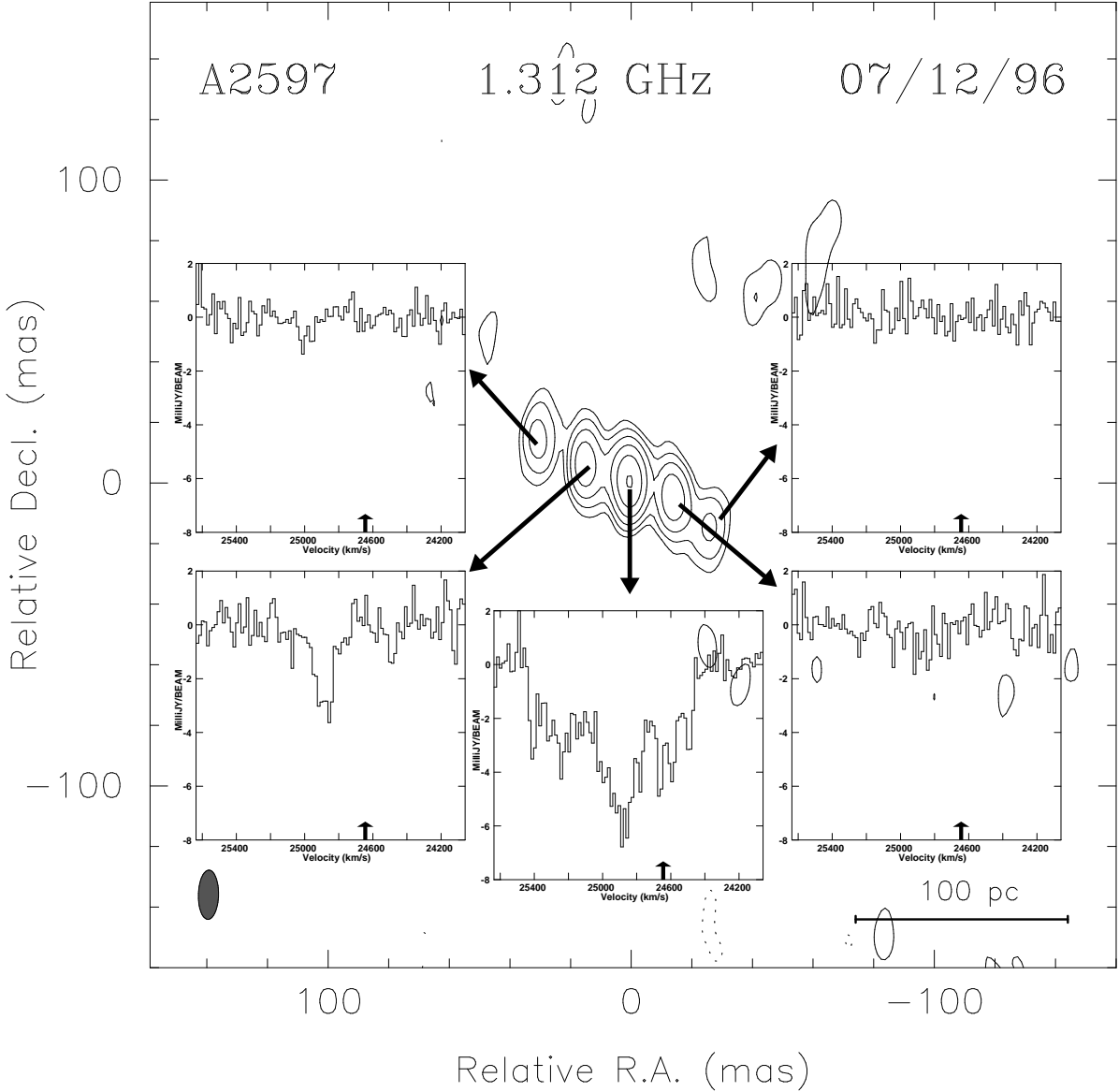}
\caption{The 1.3 GHz continuum image from Fig.~1 with the spectra
for each of the 5 components overlaid.  The velocity resolution is
7.1 km s$^{-1}$.  The arrow at 24673 km s$^{-1}$ indicates the systemic
velocity derived from the optical redshift as discussed in the text
and the width of the arrowhead represents the 1 $\sigma$ uncertainty of
100 km s$^{-1}$.
\label{fig3}}
\end{figure}

\placefigure{fig3}

\begin{figure}
\vspace{17.1cm}
\includegraphics{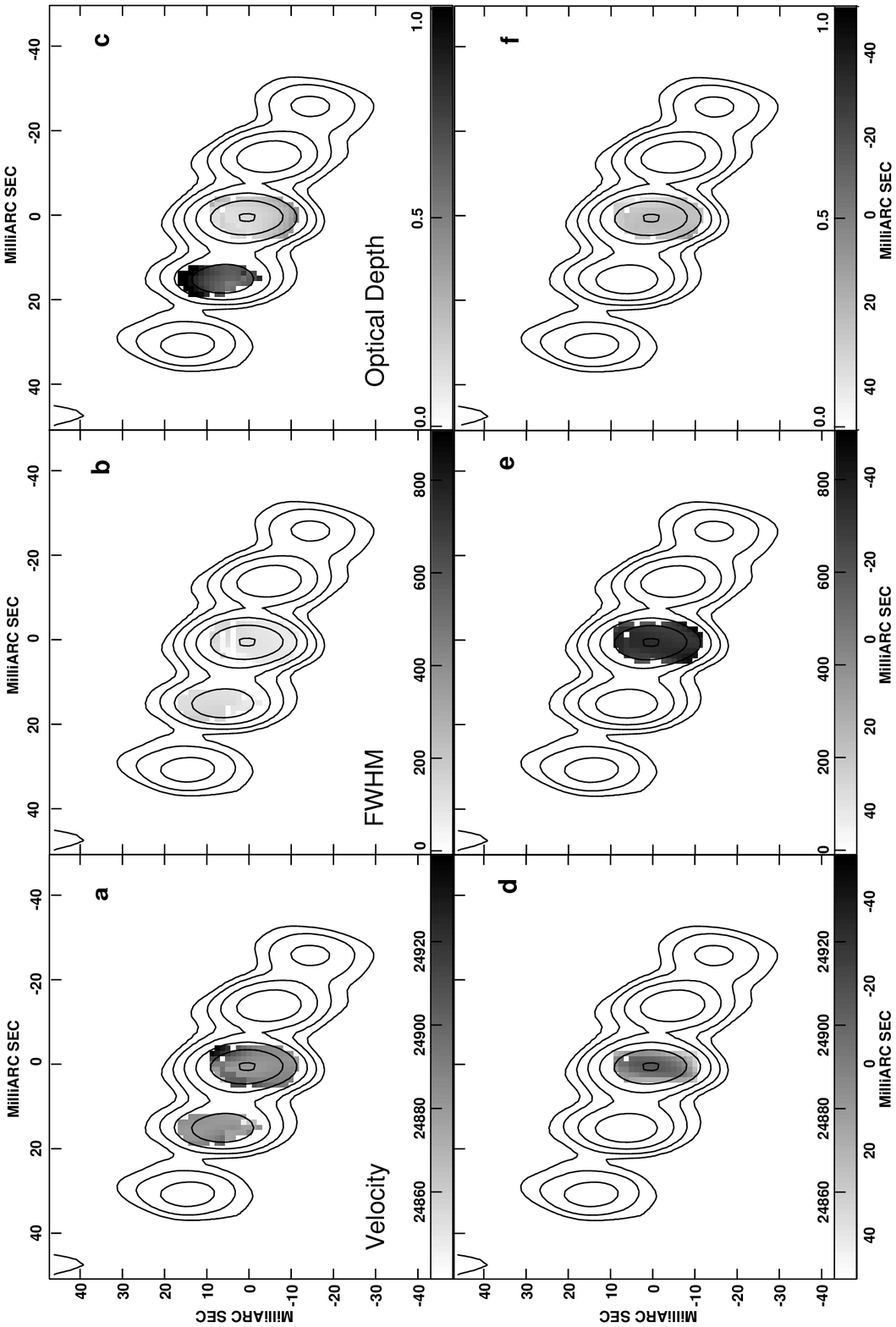}
\caption{Optical depth ($\tau$), velocity (km s$^{-1}$) and FWHM (km s$^{-1}$) grayscale plots
obtained by fitting 1 dimensional Gaussians to the \HI\ spectra pixel
by pixel.  No fit was attempted in the absence of 3 channels
with 1.5 mJy or more of absorption.
The top row is for the narrow component and the bottom row is for 
the broad component.
\label{fig4}}
\end{figure}

\placefigure{fig4}

The broad (412 km s$^{-1}$ FWHM) \HI\ line centered at 24604 $\pm$ 17
km s$^{-1}$ discovered by O'Dea \etal\ (1994) is not seen towards any
component on the parsec scale.  This is not surprising since it was
spatially resolved, and appeared strongest towards the NE lobe of
\pks.  In all likelihood the continuum and \HI\ emission from the NE
lobe has been resolved out.  The narrow (221 km s$^{-1}$ FWHM)
spatially unresolved line seen by O'Dea \etal\ at 24886 $\pm$ 5 km
s$^{-1}$ towards the nucleus is detected and spatially resolved by our
observations.

The very broad (735 km s$^{-1}$ FWHM) line was not detected by O'Dea
\etal.  Its depth of 3.8 mJy should have been detectable by them even
when diluted by unabsorbed emission within the beam, but given their
poorer velocity coverage (1304 km s$^{-1}$ after editing) it is
possible that this very broad line was lost in the continuum
subtraction.

\section{Discussion}

\subsection{Relation of Symmetric Structure to \HI\ Absorption }


The parsec scale structure of \pks\ appears extremely symmetric at
both 1.3 and 5 GHz.  If the jet components are intrinsically similar
and were ejected from the core simultaneously, then either (1) the
radio source must lie within a few degrees of the plane of the sky or
(2) the bulk motion of the jets must be nonrelativistic.  VLBI
observations of two other FR-I radio galaxies embedded in X-ray
cooling flow clusters (Hydra~A -- Taylor 1996; and 3C~338 -- Feretti
\etal\ 1993) have revealed remarkably symmetric structures on the parsec
scale whereas over 95\% of all AGN in complete flux limited samples
have one-sided core+jet morphologies (Taylor \etal\ 1996).  If even
low power radio galaxies have jets that start out relativistic 
on parsec scales (Giovannini \etal\ 1998), then the jets of all 3 sources
must be oriented with a few degrees of the plane of the sky.  The
chance of 3 out of 13 radio galaxies embedded in cooling flow clusters
(see Taylor, Barton \& Ge 1994)
being within 5 degrees of the plane of the sky is $\sim$3\%.  
Alternatively, it has
been suggested (e.g., Soker \& Sarazin 1988) that cooling flows may
decelerate the radio jets, although such disruption has been
considered to occur at the sonic radius (where the inflow velocity
exceeds the sound speed) somewhere between 0.1--10 kpc.
Unless the cooling flows have an affect on the jet closer to the
nucleus than previously thought it is difficult to understand why the
parsec scale jets in Hydra~A, 3C~338, and \pks, would have such low bulk
velocities.

Examination of the parsec scale structure of sources with \HI\ 
detected in absorption (Hydra~A -- Taylor 1996; 1413+135 -- Perlman
\etal\ 1996; N\,3894 -- Peck \& Taylor 1998; 1946+708 -- Peck, Taylor,
\& Conway 1999) indicates that a symmetric parsec-scale structure
correlates strongly with the presence of \HI, although there are
counter-examples (e.g. Mrk\,231 -- Carilli, Wrobel \& Ulvestad 1998).
3C\,84 could be an intermediate case -- observations by Vermeulen
\etal\ (1994) suggest an inclination of $\sim$45$^\circ$.  The most
likely explanation for this correlation is that symmetric sources are
oriented close to the plane of the sky such that their appearance is
not made one-sided by Doppler boosting effects.  This then provides a
favorable viewing angle of the radio source through a disk or torus of
atomic gas that is perpendicular to the jet axis.  This picture is
consistent with unified schemes, and our limited understanding of how
the central engine in AGN might be fed.

In the inner regions of the disk a large fraction of the gas must be
ionized by the central engine.  This will result in free-free
absorption of the radio continuum at frequencies below $\sim$5 GHz as
seen in 3C\,84 (Walker \etal\ 1998), Hydra~A (Taylor 1996), 1946+708
(Peck \etal\ 1999), and \pks.  Another result of the dense ionized
gas, if magnetized, could be extremely high Faraday rotation measures
(RMs).  Owing to the lack of any polarized flux in these systems, it
has not been possible to directly measure the RMs in any source with
\HI\ absorption, but the presence of very high RMs could explain
why no polarized flux is detected from these sources.

\subsection{Location of the Atomic Gas }

The sharp spatial variations in optical depth on scales of order 10 pc
(Fig.~4), and the detection of the very broad line only against the
core, argue that the atomic gas we are seeing is located in the
central 10s of parsecs of \pks, and is not just a chance encounter
with a cloud in the galaxy.  If the gas is distributed in a disk
centered on the core then it must be fairly thin ($<$20 pc) so as to
cover the core and E2, but not W2.  This situation is somewhat
different from the radio galaxy 1946+708 (Peck \etal\ 1999) where \HI\
gas is seen across the entire 60 parsec extent of the radio source.
However, in 1946+708 as in \pks, a broad component ($\sim$300 km s$^{-1}$
wide) is seen only within $\sim$10 pc of the core.  The 735 km s$^{-1}$ FWHM
line in \pks\ is, to our knowledge, the broadest \HI\ absorption line,
and is considerably wider than the optical emission lines from the
nucleus of \pks\ which have a velocity dispersion of 270 km s$^{-1}$ (Voit \&
Donahue 1997) albeit from much lower spatial resolution observations.

The fact that both lines in \pks\ are redshifted by $\sim$220 $\pm$
100 km s$^{-1}$ may imply that the gas is in-falling.  This is similar to
the trend found by van Gorkom \etal\ (1989) for 6 of 8 \HI\ absorption
systems in a sample of nearby ellipticals.  If this gas is in a disk,
then the disk is not in an orderly Keplerian rotation about the
nucleus, but has significant inward streaming motions.  From the very
broad linewidths, much broader than the thermal linewidths for any
plausible conditions, we can also infer considerable velocity
structure in the disk, probably due to turbulence.
Alternatively, given the velocity offset of the very broad line from
the absorption line redshift of 232 $\pm$ 100 km/s, it is possible
that the very broad line is at the systemic velocity.  In this case it
would be natural to explain the linewidth by rotation.  Unfortunately,
we don't know the radial extent of the disk since that is
perpendicular to the radio axis.  Assuming a radius of 10 pc, the mass
required to produce the observed linewidth is 2 $\times 10^8$
M$\odot$.

\acknowledgments 

GBT thanks STScI for hospitality
during a visit in which this paper was written.  This research has
made use of the NASA/IPAC Extragalactic Database (NED) which is
operated by the Jet Propulsion Laboratory, Caltech, under contract
with NASA.

%
%

\clearpage


\clearpage


\clearpage

\begin{references}

\reference{bla79}Blandford, R.~D., \& K\"onigl, A. 1979, ApJ, 232, 34
\reference{car98}Carilli, C.~L., Wrobel, J.~M., \& Ulvestad, J.~S.~1998,
AJ, 115, 928
\reference{con95}Conway, J.~E., \& Blanco, P.~R. 1995, ApJ, 449, L131
\reference{cra82}Crane, P.~C., van der Hulst, J.~M., \& Haschik, A.~D. 1982,
in proceedings of IAU Symposium No. 97, Extragalactic Radio Sources,
eds. D. S. Heeschen \& C. M. Wade, (Dordrecht, Reidel), 307
\reference{cot97}Cotton, W.~D., Fanti, C., Fanti, R., Dallacasa, D., 
Foley, A.~R., Schilizzi, R.~T., \& Spencer, R.~E. 1997, A\&A, 325, 479
\reference{dwa94}Dwarakanath, K.~S., van Gorkom, J.~H., \& Owen, F.~N.
1994, ApJ, 432, 469
\reference{dwa95}Dwarakanath, K.~S., Owen, F.~N., \& van Gorkom, J.~H.
1995, ApJ, 442, L1
\reference{fal98}Falcke, H., Rieke, M.~J., Rieke, G.~H., Simpson, C., 
\& Wilson, A.~S. 1998, ApJ, 494, L155
\reference{fer93}Feretti, L., Comoretto, G., Giovannini, G., 
Venturi, T., \& Wehrle, A.~E. 1993, {ApJ}, {408}, 446
\reference{gio98}Giovannini, G., Taylor, G.~B., Arbizzani, E., Bondi, M., 
Cotton, W.~D., Feretti, L., Lara, L. \& Venturi, T. 1998, in preparation
\reference{van89}van Gorkom, J.~H., Knapp, G.~R., Ekers, R.~D., Ekers, D.~D.,
Laing, R.~A., \& Polk, K.~S. 1989, AJ, 97, 708
\reference{jaf90}Jaffe, W. 1990, A\&A, 240, 254
\reference{jaf91}Jaffe, W. 1991, A\&A, 250, 67
\reference{koe98}Koekemoer, A.~M. \etal\ 1998, in preparation
\reference{lev95}Levinson, A., Laor, A., \& Vermeulen, R.C.
1995, ApJ, 448, 589
\reference{ode94}O'Dea, C.~P., Baum, S.~A., \& Gallimore, J.~F. 1994, ApJ, 436, 669
\reference{ode95}O'Dea, C.~P., Gallimore, J.~F., \& Baum, S.~A. 1995, AJ, 109, 26
\reference{pec98}Peck, A.~B., \& Taylor, G.~B. 1998, ApJ, 502, L23
\reference{pec99}Peck, A.~B., Conway, J.~E., \& Taylor, G.~B. 1999, in prep.
\reference{per96}Perlman, E.~S., Carilli, C.~L., Stocke, J.~T., \& 
Conway, J.~E. 1996, AJ, 111, 1839
\reference{sar95}Sarazin, C.~L., Burns, J.~O., Roettiger, K., \& McNamara, B.,
1995, ApJ, 447, 559
\reference{dif94}Shepherd, M.~C., Pearson, T.~J., \& Taylor, G.~B.~1994, BAAS, 26, 987
\reference{dif95}Shepherd, M.~C., Pearson, T.~J., \& Taylor, G.~B.~1995, BAAS, 27, 903
\reference{sok88}Soker, N., \& Sarazin, C.~L. 1988, {ApJ}, {327}, 66
\reference{tay94}Taylor, G.~B., Barton, E.~J., \& Ge, J.-P. 1994,
AJ, 107, 1942
\reference{tay96}Taylor, G.~B. 1996, ApJ, 470, 394
\reference{tay96b}Taylor, G.~B., Vermeulen, R.~C., Readhead, A.~C.~S., 
Pearson, T.~J., Henstock, D.~R., \& Wilkinson, P.~N.
     1996, in ``The Second Workshop on Gigahertz Peaked Spectrum and Compact
     Steep Spectrum Radio Sources'', eds. I.A.G. Snellen, R.T. Schilizzi,
     H.J.A. Rottgering, and M.N. Bremer (Leiden Observatory:Leiden), p. 263
\reference{rcv94}Vermeulen, R.~C., Readhead, A.~C.~S., \& Backer, D.~C.~1994,
ApJ, 430, L41
\reference{voi97}Voit, G.~M., \& Donahue, M. 1997, ApJ, 486, 242
\reference{wal98}Walker, R.~C., Kellermann, K.~I., Dhawan, V., Romney,
J.~D., Benson, J.~M., Vermeulen, R.C., \& Alef, W. 1998, in
{\it IAU Colloquium 164: Radio Emission from Galactic and Extragalactic
Compact Sources} eds. J. A. Zensus, G. B. Taylor and J. M. Wrobel
(PASP: San Francisco) Vol. 144, p. 133
\reference{whi91}White, D.~A., Fabian, A.~C., Johnstone, R.~M., 
Mushotzky, R.~F. \& Arnaud, K.~A. 1991, MNRAS, 252, 72


\end{references}
\end{document}